\documentclass[9pt,twocolumn,twoside]{pnas_template}
\templatetype{pnasresearcharticle}
\leadauthor{de Courson} 
\usepackage[utf8]{inputenc}

\usepackage{bbm}
\DeclareMathOperator*{\argmin}{argmin}
\usepackage{authblk}
\usepackage{hyperref}
\usepackage{csquotes}

\usepackage{multicol,lipsum}
\usepackage{xcolor}
\usepackage{mathtools}
\usepackage{subfig}
\usepackage{graphicx}

\usepackage{enumerate} 
\usepackage{enumitem}

\begin{abstract}
The ability to learn from others (social learning) is often deemed a cause of human species success. But if social learning is indeed more efficient (whether less costly or more accurate) than individual learning, it raises the question of why would anyone engage in individual information seeking, which is a necessary condition for social learning's efficacy. We propose an evolutionary model solving this paradox, provided agents (i) aim not only at information quality but also vie for audience and prestige, and (ii) do not only value accuracy but also reward originality -- allowing them to alleviate herding effects. We find that under some conditions (large enough success rate of informed agents and intermediate taste for popularity), both social learning's higher accuracy and the taste for original opinions are evolutionary-stable, within a mutually beneficial \textit{division of labour}-like equilibrium. When such conditions are not met, the system most often converges towards mutually detrimental equilibria. 
\end{abstract}

\significancestatement{Many aspects of our social lives rely on the efficiency of the wisdom of crowds effect: juries, committees, parliaments, financial markets, etc. But if aggregate information is more accurate, why would anyone engage in information seeking instead of free-riding the majority opinion, which would then eventually disconnect from reality? This paradox, in various incarnations, pervades the social science literature. We propose an evolutionary stable, mutually beneficial equilibrium where less accurate individual learners revel in increased popularity, while followers hedge themselves against group-think by heeding original voices. For intrinsically difficult issues, however, this virtuous equilibrium collapses.}

\title{Why does individual learning endure when crowds are wiser?}
\date{December 2020}
\author[1,2]{Beno\^ it de Courson}
\author[3]{L\'eo Fitouchi}
\author[2,4]{Jean-Philippe Bouchaud}
\author[$\star$,1,2,4]{Michael Benzaquen}

\affil[1]{LadHyX, UMR CNRS 7646, Ecole Polytechnique, 91128 Palaiseau Cedex, France}
\affil[2]{Chair of Econophysics \& Complex Systems, Ecole Polytechnique, 91128 Palaiseau Cedex, France}
\affil[3]{Institut Jean Nicod, D\'{e}partement d’\'{e}tudes cognitives, ENS, EHESS, PSL Research University, CNRS, Paris France}
\affil[4]{Capital Fund Management, 23-25, Rue de l'Universit\'e 75007 Paris, France}

\authorcontributions{MB, JPB and BdC designed research; BdC performed research, all the authors wrote the paper.}
\authordeclaration{The authors report no conflict of interest.}
\correspondingauthor{\textsuperscript{$\star$}To whom correspondence should be addressed: michael.benzaquen@polytechnique.edu}

\begin{document}
\maketitle
\ifthenelse{\boolean{shortarticle}}{\ifthenelse{\boolean{singlecolumn}}{\abscontentformatted}{\abscontent}}{}
\setlength{\parindent}{1em}

\dropcap{I}n 1785, Condorcet proved that aggregating predictions can be much more accurate than individuals' predictions \cite{condorcet1785essay}. As the group size increases, the average prediction can even approach perfect accuracy, provided that individual predictions are unbiased. Empirically, this phenomenon was famously illustrated by Galton's ``Vox populi'' paper \cite{galton}: when he gathered the estimates of an ox weight from 787 participants, the average opinion was just one pound away from reality.

Yet, the very fact that the aggregate prediction is better than individual ones creates a tension: why would anyone make up his/her mind independently when he/she would be better off sampling around and simply endorsing the majority opinion? A very close analogy can be found in the Grossman-Stiglitz paradox in financial economics \cite{grossmanImpossibilityInformationallyEfficient1980}: if, as often assumed, the price of an asset aggregates all available information, then there is no reason to gather information, which makes it absurd that the price reflects anything to begin with. 

\begin{figure}
    \centering
    \includegraphics[width=1\columnwidth]{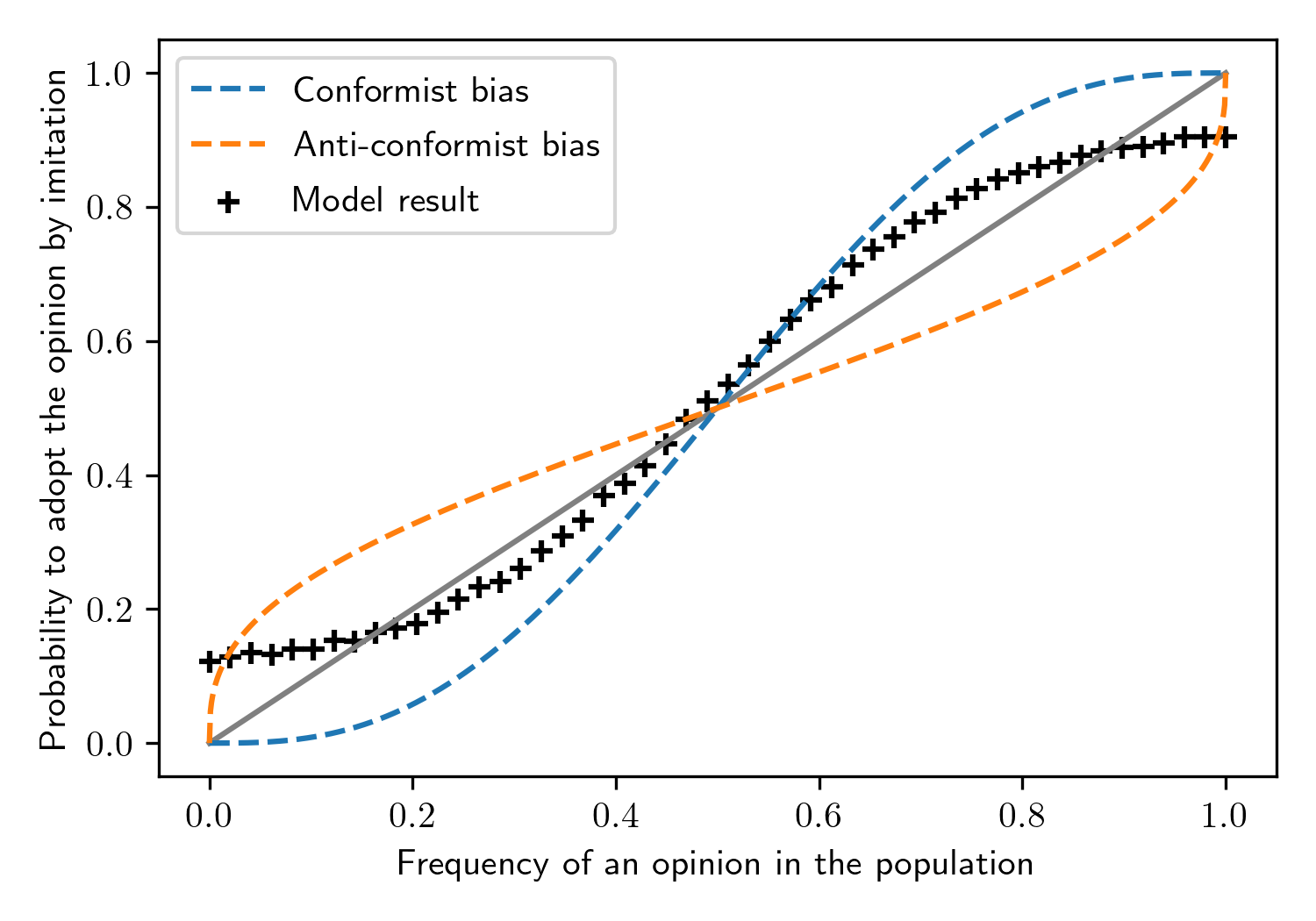}
    \caption{Definition of a (anti-) conformist bias (dashed orange and blue lines) in the cultural evolution literature~\cite{boydCultureEvolutionaryProcess1985,henrichEvolutionConformistTransmission1998}, compared with the outcome of our model (explained in \ref{sec:div}) where some endogenous heterogeneities are generated. }
    \label{fig:bias}
\end{figure}

From an evolutionary game theoretic perspective, this implies that a population where all agents decide independently would swiftly be invaded by a ``conformist'' strategy, as long as the sample size is greater than one, or if there is some information cost. This creates what Boyd, Richerson and Henrich \cite{boydCultureEvolutionaryProcess1985,henrichEvolutionConformistTransmission1998} called a conformist bias (technically defined by the fact that the probability to adopt the belief held by the majority is greater than the frequency of the majority opinion in the population, see Fig.~\ref{fig:bias}). Importantly, this idea has been invoked to explain group-level cultural differences: as Henrich~\cite{henrichCulturalGroupSelection2004a} puts it, {\it without a conformist component to create ‘cultural clumps’, social learning models predict (incorrectly) that populations should be a smear of ideas, beliefs, values and behaviours, and that group differences should only reflect local environmental differences}.

This rationale brings however an internal tension within Condorcet's jury theorem, which rests on a crucial assumption: that opinion are somewhat independent in the probabilistic sense. When allowing imitation, the theorem jeopardises the very independence it is assuming: if most of the population imitates others,  opinions are no longer independent, but converge towards one another (an effect that has been found empirically~\cite{lorenzHowSocialInfluence2011}). This independence breakdown has consequences -- at some point, opinions are so tightly correlated that the accuracy achieved by imitation collapses, as individuals agree more with one another than with reality. The snake bites its own tail: the increase of accuracy achieved by sampling the opinion of others triggers a spread of imitators, which continues until all anchor to ``truth'' has disappeared.

This intuition, elegantly modeled by Curty \& Marsili~\cite{curtyPhaseCoexistenceForecasting2006} (see Section \ref{sec:2}), sheds light on herding phenomena. However, it leads to paradoxical predictions. In particular, at equilibrium, 
\begin{itemize}
  \setlength\itemsep{0.2em}
\item  the vast majority of agents imitate, so that virtually every agent holds the same belief (i.e. full in-group cultural homogeneity, which is also a prediction of the conformist transmission literature~\cite{henrichEvolutionConformistTransmission1998}). This conclusion can be understood from Fig.~\ref{fig:bias}: if we think of imitation as a dynamic process, then frequencies $0$ and $1$ are the only stable fixed points of the system.
\item  hence, the ``wisdom of crowds'' cannot work, since the aggregation of similar opinion does not lead to more accurate predictions. This result is known as the ``Rogers' paradox''~\cite{rogersDoesBiologyConstrain1988} in the special case where agents only imitate a single person (see Section~\ref{rogers}).
\end{itemize} 

Both conclusions are unrealistic. First, there are countless examples of the wisdom of crowds in action \cite{surowieckiWisdomCrowdsWhy2004,senningerWisdomCrowdExperiments2018,kurversBoostingMedicalDiagnostics2016}, including evidence that access to other people's opinion 
improves collective accuracy \cite{lorenzHowSocialInfluence2011,novaestumpIndividualsFailReap2018,wolfAccurateDecisionsUncertain2013,clementInformationTransmissionMovement2015}.

Second, there is usually a substantial in-group heterogeneity in people's beliefs (see the conclusion of Ref. \cite{erikssonArePeopleReally2009}). A study trying to measure within- and between-group cultural variation incidentally found the first component to be the larger than the second  \cite{bellCultureRatherGenes2009}. Furthermore, humans are found to use individual learning quite often \cite{effersonConformistsMavericksEmpirics2008,erikssonBiasesAcquiringInformation2009,mesoudiExperimentalComparisonHuman2011} and to under-use social information \cite{novaestumpIndividualsFailReap2018,morin2020social}, a phenomenon known as ``egocentric discounting'' \cite{yanivAdviceTakingDecision2000}, which leads to a sub-optimal situation in terms of belief accuracy, both at the individual \cite{novaestumpIndividualsFailReap2018,mesoudiExperimentalComparisonHuman2011} and collective \cite{miuCulturalAdaptationMaximised2020} levels. Also, the very tendency to conformism is disputed \cite{erikssonArePeopleReally2009}: minorities seem to have a literally disproportionate social influence, greater than their share in the population \cite{lataneSocialImpactMajorities1981}. In the famous Asch experiment \cite{aschStudiesIndependenceConformity1956}, often heralded as proof of human conformism, the proportion of participants following the majority (false) opinion collapsed when the experimenter instructed one of his accomplices to hold the correct opinion. 

Anyway, as Eriksson \& Coultas \cite{erikssonArePeopleReally2009} argue in response to Henrich's aforementioned statement, {\it we can turn the argument around: with a conformist component, social learning models predict, incorrectly, that populations should be culturally homogeneous. In this light, it is not surprising that the evidence from social psychology does not support the Conformist Hypothesis.} However, an anti-conformist bias does not seem credible either: it would imply evenly split populations (in Fig. \ref{fig:bias}, $1/2$ is the only stable fixed point of the orange curve), which is unrealistic.

Hence, we need a model which, while allowing individuals to exploit the wisdom of crowds, accounts for its effectiveness at equilibrium and for persistent within-group heterogeneity. These two phenomena are actually intertwined. The wisdom of crowds effect is contingent on some opinion diversity in the group, otherwise the opinion of the crowd is the same as that of any individual. In a theoretical model, Scott E. Page indeed showed that {\it the amount by which the crowd out-predicts its average member increases as the crowd becomes more diverse} \cite{pageDifferenceHowPower2008}. In other words, in order to benefit from the wisdom of crowds effect, one should aggregate {\it diverse} opinions.

This is a feature that should be  incorporated in theoretical models. Previous studies implicitly assume that agents listen to randomly chosen other agents. Instead, we propose that agents can develop a ``taste for originality'', implemented as a preferential attachment to individuals displaying deviant opinions, in a dynamic network representing who listens to whom. This taste for originality is not only normatively warranted, but also empirically reasonable. Several studies have found that individuals having atypical opinions were rated as more likeable, courageous or admirable \cite{dyneNaturalisticMinorityInfluence1996,jettenDevianceDissentGroups2014,nemethModellingCourageRole1988}. Other studies have found that individuals put more weight on advice from people who form their opinions independently \cite{bloomfieldExperimentalInvestigationPositive2009,hessPsychologicalAdaptationsAssessing2006,mercierUtilizingSimpleCues2019} (see \cite{mercierMajorityRulesHow2019} for a review). Similarly, financial analysts who produce ``bold forecasts'' tend to be preferentially heeded \cite{hong}. Agents seem to be intuitively aware of the danger of blindly following the ``crowd''.  

Being independent-minded 
generates {\it interesting} -- because unexpected -- opinions, likely to be listened to. This could be the primary goal of certain individuals. In current evolutionary models, fitness is proxied by belief accuracy, which means that humans are assumed to be accuracy-maximizers. Here, we propose that humans also care about their social influence (which is empirically related to reproductive success \cite{vonruedenWhyMenSeek2011}). The dynamic network representing who listens to whom allows one to do so in a simple way: agents simply maximize a combination of their opinion accuracy and the number of people who listen to them. 

To sum up, we introduce two novel assumptions: (i) agents have an incentive to be listened to, and (ii) agents can choose to listen to deviant opinions in order to benefit from the diversity effect. This framework allows us to shed a new light on individual learning: although it is sub-optimal in terms of accuracy, it is a way to signal independent mindedness, which makes one worthy to be listened to. It leads to a kind of division of labour equilibrium: as they do not benefit from the wisdom of crowds effect, individual learners achieve lower accuracy, which is offset by the reward they get from being listened to. Such reward may be prestige, influence, or more concretely university positions, consulting fees, copyright, etc.
Our scenario is consistent with the empirical findings we mentioned: (i) the effectiveness of the wisdom of crowds effect despite widespread social learning, (ii) a stable non-zero degree of within-group cultural diversity, and (iii) an over-reliance on individual learning, with respect to what accuracy-maximisation would require. 

The paper is organised as follows. In Section~\ref{sec:2} we recall the seminal model of Curty \& Marsili, which serves as groundwork for our paper. Then, we proceed by making the model gradually more complex. In \ref{sec:net}, we introduce an adaptive network of imitation and assess the consequences on the model's predictions. In \ref{sec:orig}, we relax the assumption that agents imitate randomly, and we allow them to listen preferentially to original opinions. In \ref{sec:evo}, we introduce an evolutionary framework to endogenize the strategy (individual or social learning) and the taste for originality. In \ref{sec:anti-conformists}, we explore the possibility of an anti-conformist bias to ``mimic originality''.

\newpage 

\tableofcontents

\section*{Methods and results}
\addcontentsline{toc}{section}{Methods and results}
\subsection{The Curty-Marsili forecasting game}
\label{sec:2}

\subsubsection{The classical setting}

The seminal model of Curty \& Marsili~\cite{curtyPhaseCoexistenceForecasting2006} considers a population of agents making a binary forecast for the next time step. Agents have two possible strategies: looking for information ($I$) or being ``follower'' ($F$) and relying on others. An agent playing the $I$ strategy has a probability $p > 1/2$ of being right (a quantity that we will, henceforth, call the ``accuracy'' of the strategy). The follower strategy consists in an iterative imitation process, specified as follows:
\begin{itemize}
 \setlength\itemsep{0.2em}
    \item All agents playing $F$ randomly initialize their opinion, with a $1/2$ probability of being right: they have no private information, and thus a neutral prior.
    \item One of them, randomly chosen, observes a random sample of $m$  individuals, and endorses the majority opinion among them (for simplicity, $m$ is assumed to be odd to avoid ties).
    \item The previous step is repeated a large number of times until the proportion of followers holding the right opinion converges to a value $\hat{q}$.
\end{itemize}

\begin{figure}
    \centering
    \includegraphics[width =1\columnwidth]{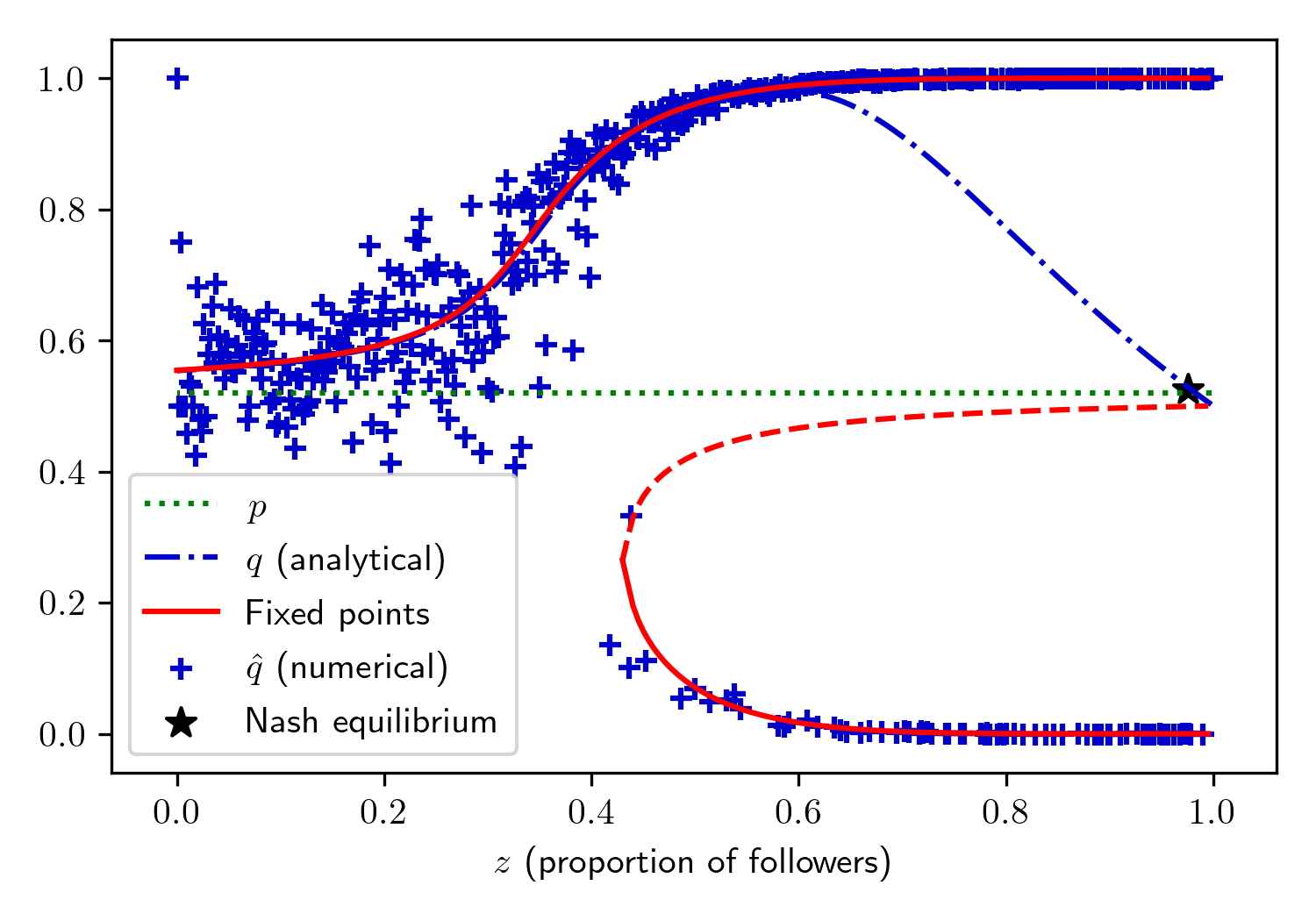}
    \caption{Impact of followers prevalence $z$ on their accuracy $q$, for $m=11$, $N=2000$, and $p=.52$ }
    \label{fig:1}
\end{figure}

Unlike the $I$ strategy, the $F$ accuracy depends on the fraction $z$ of followers in a non-trivial way, see Fig.~\ref{fig:1}. For a small fraction $z$, there is only one attractive fixed point $q_>$, larger than $p$, which means that herding is efficient. This is simply the Condorcet theorem mechanism at play: aggregating information yields more accurate predictions. Furthermore, the performance of followers increases at first with their number, as  the probability to sample from an accurate follower increases, thereby improving sample quality. Notwithstanding, above a critical value $z_c$, the system becomes bi-stable: $\hat{q}$ has two attractive fixed points (solid red line), and an unstable one in the middle (dashed red line), which marks the frontier between the two basins of attraction. Thus, the population has two possible end states, one where most people are right, and another where most people are wrong.

More intuitively, there is for $z > z_c$ a substantial probability (increasing with $z$) that, due to initial conditions, followers trap themselves in the wrong self-reinforcing forecast. When $z$ becomes very large, these two fixed point are close to $0$ and $1$, the followers basically reach a consensus. This is a herding phenomenon: when imitation is too frequent, followers are mostly imitating one another and no longer aggregating genuine information. It is possible to compute analytically $q = \mathbb{E}[\hat{q}]$ \cite{curtyPhaseCoexistenceForecasting2006}, which gives the bell-shaped blue dotted curve shown in Fig.~\ref{fig:1}.

Under the hood, this is due to the fact that the more numerous the followers, the less independent their opinions. As we mentioned in the Introduction, the super-accuracy of followers relies on independence: if all individuals agree, aggregating opinions has no effect. Worse still: as followers have no information of their own but rely on others to make up their mind, a population made of 100\% followers would decide purely by chance, which means that $q$ has to fall towards $1/2$ as $z\to 1$. 
Furthermore, there is a point $z^{\dagger}\lesssim 1$ (i.e. the vast majority of agents are followers) where $q = p$. If the agents are rational accuracy-maximizers, this point is the only Nash equilibrium.

The message of the model is twofold. First, the equilibrium of the game is the point where, even though most people learn from others, learning offers no longer any gain of accuracy ($p = q$, reminiscent of Rogers' paradox \cite{rogersDoesBiologyConstrain1988} -- see Section~\ref{rogers} in the Discussion for more details). In other words, imitation behaviour spreads until the snake bites his own tail. There is ``consensus'' -- nearly all the population adopts either the right or the wrong forecast. This model can thus explain, e.g., herding effects among financial analysts \cite{guedjExpertsEarningForecasts2004}.

This model shares the basic features and conclusions of the conformist transmission literature. In Fig.~\ref{fig:bias}, the dotted-blue curve represents the probability that an imitator acquires the right opinion by using the $F$ strategy, depending on its prevalence in the population. As we already noted, this shape matches Boyd \& Richerson's definition \cite{boydCultureEvolutionaryProcess1985} of having conformist bias (the probability to adopt the belief held by the majority is greater than the frequency of the majority opinion in the population). 

\subsection{An adaptive social network}
\label{sec:net}
\subsubsection{Assumptions}

One of the limitations of the Curty-Marsili framework is the assumption that followers sample randomly their consociates to make up their mind, thereby neglecting all possible effects of reputation, trust, and epistemic links between agents. In order to represent the fact that agents often rely on a network of people they know and trust, we embed the game in a dynamic directed graph, where each agent $i$ is a node. A link $i \rightarrow j$ signifies that $i$ listens to $j$ when he/she makes up his/her mind. The network follows the following rules.
\begin{itemize}
\setlength\itemsep{0.2em}
    \item Each agent $i$ initializes its network $\Omega_i$ of nearest neighbours by picking uniformly $m$ agents in the population.
    
    \item The members $j \in \Omega_i$ are endowed with $i$-dependent scores $S_i^j$, initialized at $0$. They are updated following:
    \begin{eqnarray}
    \forall j \in \Omega_i,\quad S_i^j(t+1) = (1 - \gamma)S_i^j(t) + R^j(t),
        \end{eqnarray}
    with $0 < \gamma \leq 1$ a memory decay rate, and a reward $R^j(t) = 1$ if $j$ was right at time $t$, $0$ otherwise. 
    
    \item At each time step, $i$ drops the lowest performer $j^\star=\argmin(S_i^j,j \in \Omega_i)$ with probability $\gamma$ (which controls the speed of the network evolution), and picks a substitute. The probability for a ``target'' agent $j^{\text{target}}$ to be chosen as a substitute is proportional to $a + k(j^{\text{target}})$, with $a > 0$ a small incompressible weight that avoids diversity depletion (when $a = 0$ an agent who has lost all its audience has no possibility to ever re-enter the network), and $k(j^{\text{target}})$ is $j^{\text{target}}$'s in-degree (measuring his/her ``reputation''). This picking process excludes oneself and all agents already present in $\Omega_i$.
\end{itemize}

For the moment, we keep the strategies ($I$ and $F$) fixed and exogenous, and observe the phenomenology of the model; we shall in a second stage let the strategies evolve depending on their respective payoffs. Henceforth, we choose  $z$ (the fraction of agents playing $F$) close to 1, and $p=0.52$. 

\subsubsection{Collapse of the informed audience share}

With this modification of the Curty-Marsili model we find that, perhaps surprisingly, $q$ now drops below $p$, and settles around $1/2$ (Fig.~\ref{fig:tryptic}(c)).

This is caused by a subtle phenomenon. In the network, the scores $S_i^j$ are supposed to estimate the accuracy of an agent. So, we should expect that when $q < p$, the followers tend to fill their network with $I$ agents, who are on average more accurate. Paradoxically, the opposite happens. Even though the scores of the informed agents are on average higher than those of the followers, the latter have a much lower variance (see Fig.~\ref{fig:tryptic}(a)). Indeed, as followers tend to make the same prediction, their scores fluctuate less than those of $I$ agents who make independent predictions. As the network is updated by withdrawing worst performing agents it is often, counter-intuitively, an $I$ agent who is jettisoned. Hence a collapse of the informed audience share, defined as the proportion of links directed towards an $I$ agent.

Therefore, the systems ends up in a situation where the tiny minority of agents who have genuine information are not heeded. In our model, as often in real life, being the only one to be wrong has much stronger consequences than being the only one to be right. Being wrong means being wiped out the network, while being right does not provide any particular advantage in terms of audience share. The informed agents suffer most from this asymmetry. This echoes some empirical results: Yaniv \& Kleinburger \cite{yanivAdviceTakingDecision2000} found that it was {\it easier for advisors to lose a good reputation than to gain one}. This is also reminiscent of Kahneman \& Tversky's prospect theory \cite{kahnemanProspectTheoryAnalysis2012}: if average performance is the point of reference, then being the only one to be wrong is a loss, which is psychologically over-weighted compared with being the only one to be right. Anecdotal evidence suggests that this also true for asset managers, who get badly punished when they are alone in a drawdown.

\subsubsection{Broad in-degree distribution}
\label{sec:power_law}

\begin{figure}
    \centering
    \includegraphics[width = 1\columnwidth]{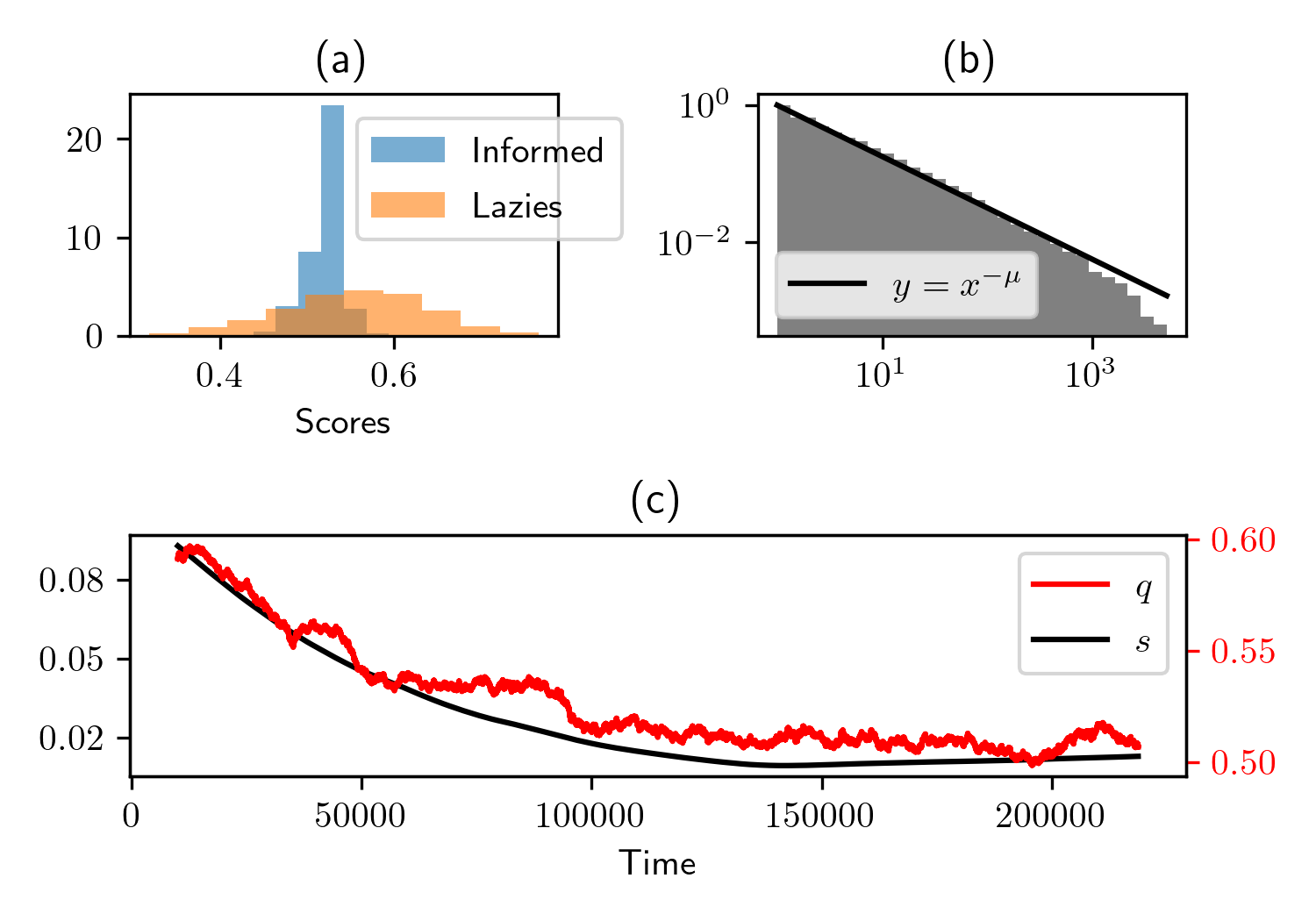}
    \caption{(a) Distribution of scores, (b) complementary {\sc cdf} of in-degrees $k$, and (c) dynamics of the model, all for $z=.98$, $p=.52$. Panel (b) reveals that the distribution of in-degrees decays as a power-law $k^{-1-\mu}$ with $\mu \approx .75$. A value of $\mu \leq 1$ means that the first moment of $k$ formally diverges, symptomatic of a ``winner-takes-all'' effect.}
    \label{fig:tryptic}
\end{figure}

In our network, the out-degrees (the number of agents a given agent is listening to) are all fixed to $m$. The dynamics of in-degrees (the number of agents listening to someone) evolves dynamically. Quite interestingly, we find that the in-degrees distribution develops a power law tail, see Fig.~\ref{fig:tryptic}(b). This can be interpreted as the spontaneous emergence of ``opinion leaders'', i.e. agents whose opinion has a systemic impact on the population. This scale-free topology is due to the assumption that the probability to pick a given node is an affine function of its in-degree. Nodes with in-degree $k \gg a$ therefore grow exponentially with time. On the other hand, the probability that the score of an ``opinion leader'' remains by chance above a certain threshold that shields him/her from losing some followers decays exponentially with time. This ``battle of exponentials'' naturally leads to power-law distribution for $k$, with an exponent that depends on the parameters of the model (see e.g. \cite{more_levy}). 

\subsection{The value of originality}
\label{sec:orig}

The results of the previous section are somewhat unsatisfying. In the situation where $q$ is close to $1/2$, any follower would benefit from listening to an informed instead of a follower. This is an assortment failure, and the situation described above is only an equilibrium because our assumptions prevent efficient assortment. 

While it makes sense to assume that the strategies themselves ($I$ or $F$) are not directly observable, opinions are public and may be used to infer the underlying strategies. But as we repeatedly pointed out, at equilibrium informed agents cannot be identified by their accuracy (in the Curty \& Marsili's equilibrium, $p=q$). 

Instead, the most distinctive feature of $I$ agents is their originality. Since their forecast is not influenced by anyone, the correlation between their opinion and the majority opinion is zero (or slightly above zero if the agent has a substantial influence in the network), whereas this quantity is close to one for followers. This distinctive feature can plausibly be observed by followers, and be used to identify potentially informed sources and listen preferentially to them. More mundanely, agents with original opinions are often deemed competent (they are said to ``think outside the box'', or hold ``bold'' opinions \cite{hong}).

This intuition can be formalized by introducing a proxy of originality, which we define to be the distance observed by agent $i$ between the forecast of an agent $j$ in $\Omega_i$ and the average forecast in $\Omega_i$. Such distance quantifies how deviant a given forecaster is, from the point of view of a follower. Including it in the scores update rule in the simplest way amounts to writing:
    \begin{eqnarray} \nonumber
S_i^j(t+1) &=& (1 - \gamma)S_i^j(t) + R^j(t) + \alpha \big\vert R^j(t) -  \overline{R}_i(t) \big\vert \\   \overline{R}_i(t) &:=& \textstyle \frac1m  \sum_{k = 1}^m{
R^k(t)}
    \label{eq:update_rule}
    \end{eqnarray}
where $\alpha$ is a weighting parameter, measuring by how much  agents value {\it originality} over {\it  accuracy}. 

In information-theoretic terms, given the opinion of an original agent, the opinion of another randomly drawn unoriginal one has a null Shannon entropy. Conversely, the opinion of an original agent has a constant positive Shannon entropy, irrespective of the opinions that have been already voiced, since the random variable representing his/her opinion is independent from all other opinions. 
In other words, listening to an original agent is a way to diversify information sources, whereas listening to a consensus-follower is a waste of time. Hence, listening to original agents can actually be seen as a form of protection against group-think, which agents are arguably wary of. 
We start by fixing $\alpha$ as an exogenous parameter before letting it evolve endogenously to an equilibrium value. 
\begin{figure}[t!]
    \centering
    \includegraphics[width=\columnwidth]{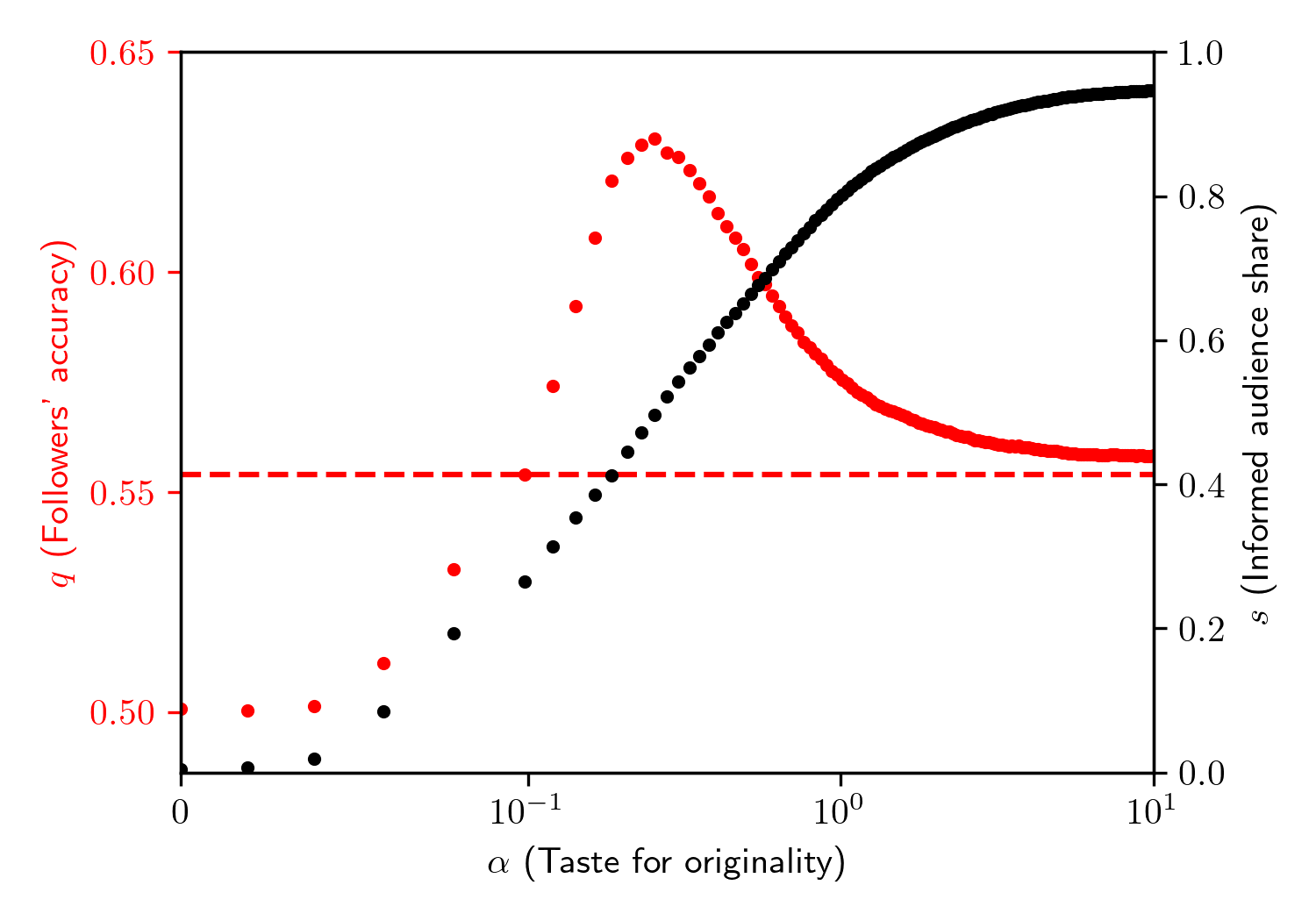}
    \caption{Impact of $\alpha$ on $I$ audience share $s$ and followers accuracy $q$ (with the asymptotic value of $q$ as a dashed line when $s \rightarrow 1$)}
    \label{fig:3}
\end{figure}

Figure~\ref{fig:3} reveals that the effect of the originality term in~\eqref{eq:update_rule} is twofold. First, as $\alpha$ increases, predictably, the taste for originality increases the audience of informed agents in the network (red markers, left axis). This means that our modified score update rule is successful in implementing the idea of beneficial originality. Perhaps surprisingly though, the effect of $\alpha$ on $q$ is {\it non-monotonic}. As $\alpha$ increases further, $q$ reaches a maximum beyond which it decays.

In order to understand why this is the case, recall that in Fig.~\ref{fig:1}, $z$ was found to have a bell-shaped effect on $q$. In the present framework, the informed audience share $s$ in the network (black markers, right axis) somehow plays the role of $z$ in Curty \& Marsili's model. There needs to be an optimal fraction of informed agents in the network in order to maximize the wisdom of crowds. It is due to the fact that when $q > p$, followers are better forecasters than informed, leading to a mean-variance trade-off between diversifying information sources and listening to good forecasters.

Finally, increasing $\alpha$ concentrates the audience on the very small number of informed agents. Thus, we expect that some $I$ agents have massive in-degrees, but numerical results are, again, counter-intuitive. The higher the $\alpha$, the lower the maximum in-degrees (not shown). This makes sense if one accounts for the fact that informed draw their high audience share from their originality. Hence, as soon as an informed agent is in the network of a significant share of the population, it acquires a systemic impact which undermines its very originality. The effect of originality on scores therefore leads, self-consistently, to a limitation of the audience share of opinion leaders.


\subsection{Preliminary summary and open issues}
\label{sec:sum}

Our exploratory results so far have shown that there exists an optimal positive value of the taste for originality $\alpha$. One would however prefer that individuals endogenously develop such a feature. But this feature actually generates externalities: in a population with a large $\alpha$ (say, where followers listen only to informed), an agent who sets its personal $\alpha$ to $0$ in order to listen mostly to followers would have a better accuracy than others. We might thus still be in a producer-scrounger problem, where free-riding would destroy the additional accuracy brought by the taste for originality.

In addition, we have so far worked with an exogenous proportion of informed agents. But if agents try to maximize their accuracy by changing strategies, this proportion should spontaneously decrease as soon as $q > p$. The foregone conclusion is that imitation produces no accuracy gain ($p = q$ is a necessary condition to have a Nash equilibrium) or, worse, that informed agents will go extinct (which implies $q=1/2$). 

Within our framework, the only way to produce a situation where $q > p$ is to modify the objective of our agents.\footnote{In some contexts, like financial markets, another realistic possibility is that informed agents can benefit from time priority. Take that the profit of an informed agent is given by $\beta p - c$ where $\beta > 1$ measures time priority and $c$ are information costs, and that the profit of followers is simply $q$. Then one can have $q > p$ at equilibrium provided $(\beta -1) p > c$.} Concretely, if agents compete not only for accuracy, but also for in-degrees (i.e. for audience), then we might establish an efficient ``division of labour''. The lower accuracy of the informed would be compensated by their large in-degrees, due to the taste for originality that followers develop.

Thus, neither the taste for originality nor the survival of informed agents can be taken for granted at this point. We need to investigate these issues in a dynamical setting, i.e. by making both $z$ and $\alpha$ endogenous. The next section introduces an evolutionary framework precisely to address this issue.

\subsection{An evolutionary framework}
\label{sec:evo}
At this point, analytical computations of an evolutionary equilibrium appear out of reach. 
We thus choose a dynamical mutation-selection algorithm, which allows us to find the equilibrium numerically, and a possible description of how the psychological traits we study evolved biologically. To do so, we need two ingredients: genes (the endogenous part of the model) and a fitness formula (maximized by evolution).

\subsubsection{The genotype}
We assume that agents have two genes:
\begin{itemize}
\setlength\itemsep{0.2em}
    \item The strategy gene, which has two alleles: informed ($I$) and follower ($F$)
    \item The originality gene, with also two alleles: $\alpha = 0$ or $\alpha = \alpha^\star >0$. 
\end{itemize}
Note that henceforth the taste for originality is no longer continuous, but instead binary ($\alpha=0$ or $\alpha^\star$), which helps for interpreting the results. Strictly speaking, what is endogenous here is not the value of $\alpha$ but rather the fraction of agents having $\alpha = \alpha^\star> 0$. 

Three phenotypes can then be distinguished: 
\begin{itemize}
\setlength\itemsep{0.2em}
    \item Followers preferring originality, thereafter called ``dandies''\footnote{We chose this word to signify that these agents also imitate, but do so in a way that make them deviant}
    \item Followers without this taste, or ``conformists''
    \item Informed (for them, the second gene makes no difference, as they do not use their social network)
\end{itemize}

The natural selection dynamics is implemented in a Wright-Fisher fashion \cite{imhof2006evolutionary}. At each time step, a uniformly drawn individual is killed, and replaced by a clone of some agent chosen with a probability proportional to fitness, defined below. At each time step, each gene has also a small probability $\sigma$ to mutate to the alternative allele (typically, $\sigma=10^{-8}$ in the simulations).

\subsubsection{Costs and benefits}
In addition to the obvious aim of agents, which is to make the right forecast as often as possible, we add two further sources of costs and benefits. First, information seeking has a cost. Second, we assume that being listened to has benefits. It can represent an analyst fee, university wage or simply the prestige or the power reaped from being influential. 

Formally, we introduce the fitness of an individual $F_i(t)$, which is a weighted sum of three ingredients:
\begin{itemize}
\setlength\itemsep{0.2em}
\item A measure of an agent's forecast performance, $S_i(t)$, recorded with an exponential decay memory kernel (i.e. $S_i(t) = (1-{\gamma})S_i(t-1) + R_i(t)$, with ${\gamma}$ the memory decay rate and $R_i(t)=1$ if $i$ was right at time $t$ and $0$ otherwise.
\item The cost of information ($c\geqslant 0$), incurred by the informed 
\item A measure of an agent's audience/prestige proxied by its in-degree $k_i:=\#(j/i\in \Omega_j)$, weighted by a coefficient $\omega$ quantifying how rewarding it is to be listened to, relatively to being right.
\end{itemize}
The fitness formula, in its simplest linear form, thus writes:
\begin{eqnarray}
F_i(t) = S_i(t) - c\cdot \mathbbm{1}_{i\in {I}} + \omega \cdot k_i.
\label{eq:fitness}
\end{eqnarray}

\subsubsection{A mutually beneficial division of labour}
\label{sec:results_1}

\begin{figure}
    \centering
    \includegraphics[width = 1\columnwidth]{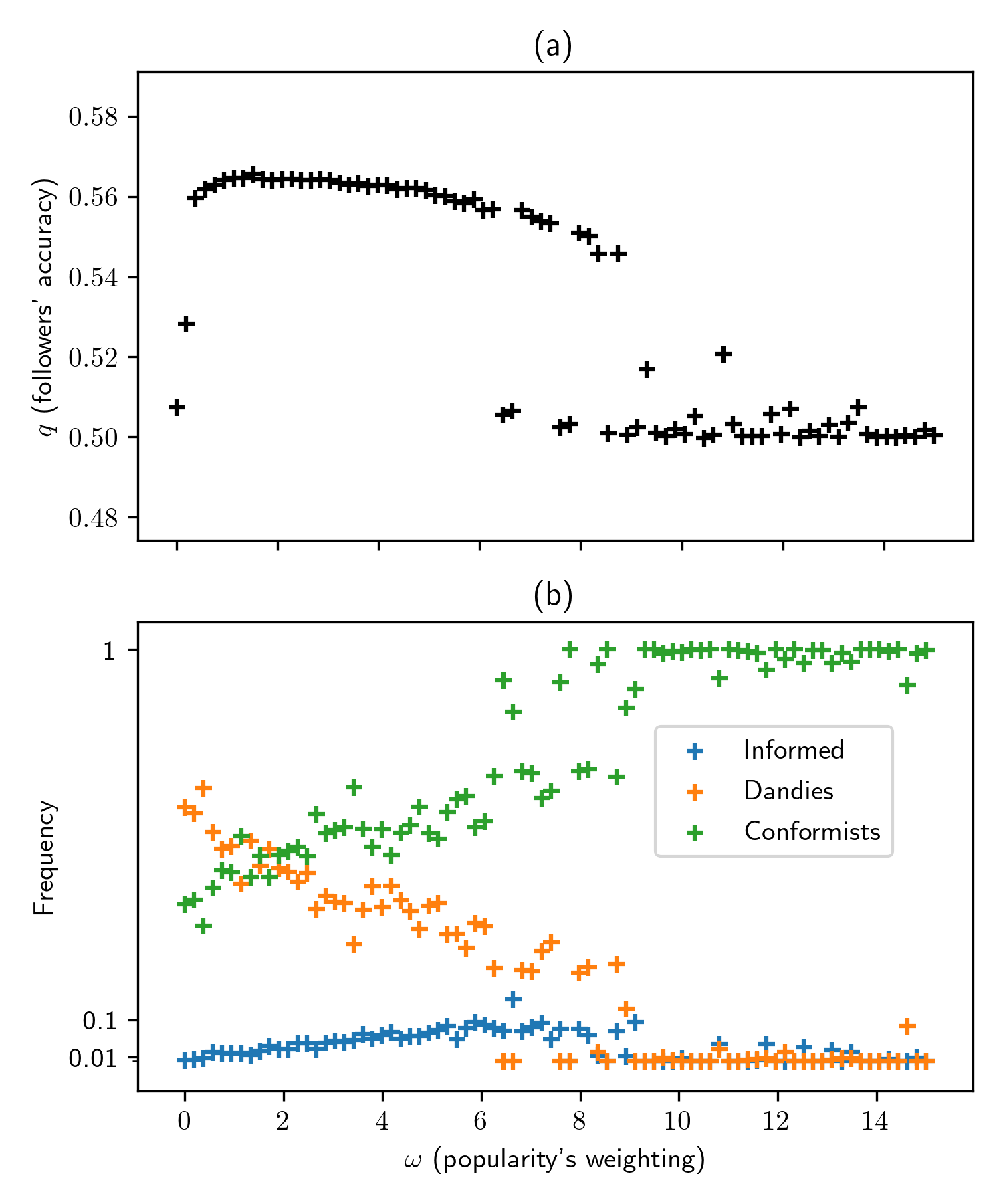}
    \caption{Effect of $\omega$ on accuracy $q$ and phenotype prevalence, with $p$=.52, c=.01 }
    \label{fig:types}
\end{figure}

Figure \ref{fig:types}(a) shows the average accuracy attained by followers as a function of $\omega$. Quite remarkably, we find that $q$ increases rapidly with $\omega$ for small $\omega$, reaches a maximum and then decreases. For intermediate values of $\omega$ the system appears to be bi-stable, with two equilibria: $q_1 \approx 1/2$ and $q_2 > 1/2$ one, the latter becoming gradually unreachable as $\omega$ increases.

In order explain this result, we plot the frequencies of types in the population (Fig.~\ref{fig:types}(b)). We see that the higher the $\omega$, the rarer ``dandies'' are, whereas the effect on the fraction of informed agents is bell-shaped. The first observation comes from the phenomenon discussed in Section~\ref{sec:net} and illustrated in Fig.~\ref{fig:tryptic}(a). Like the informed (but to a lesser extent), dandies decouple their opinion from the rest of the population. Their scores in the network are therefore more dispersed. Thus, exactly as for the informed, they tend to have fewer in-degrees than conformists. Hence, the higher the $\omega$, the  more natural selection will disfavour the dandy allele.  

The bell-shaped effect on informed agents then follows naturally. We suggested in Section~\ref{sec:sum} that the survival of informed agents was contingent on two features: the reward brought by in-degrees (that compensates their lower accuracy) and the presence of dandies (who constitute the lion's share of the audience of informed agents). Here, these conditions translate into an $\omega$ value high enough for agents to have an incentive to be followers, but not too high -- otherwise, for the reason stated in the previous paragraph, dandies go extinct, preventing informed agents from having an audience. 

To sum up, we find three distinct regimes:
\begin{itemize}
\setlength\itemsep{0.2em}
    \item Small $\omega$: informed go extinct as they have no reason not to; they are less accurate than dandies, and accuracy is the main asset. Their extinction causes $q$ to fall to $1/2$, as no one in the network has any information any longer.
    \item High $\omega$:  dandies cannot spread, as they lack in-degrees (which is here crucial). Their absence causes informed agents to go extinct, as the high reward for in-degrees does not favour them anymore.
    \item Intermediate $\omega$ (from $1$ to $5$ in Fig.~\ref{fig:types}(a)): the three types can coexist, with increased accuracy at the global level ($q > p$). 
\end{itemize}

The last regime is especially interesting and is that which we are looking for. We can describe it as a mutually beneficial ``division of labour'': informed agents do the hard work and collect reliable information, dandies listen to them to form more accurate opinions (meta-analysis), whereas conformists listen to a mix of conformists and dandies (so dandies achieve an indirect transmission of information between informed and conformists, otherwise impossible). The taste for originality is now stable, as the lower in-degrees of dandies is offset by their higher accuracy. The presence of an informed minority is also stable, as their lower accuracy is offset by their high in-degrees. 

\subsubsection{Opinion diversity}
\label{sec:div}

One can also analyse these results in terms of opinion diversity. Such diversity can be thought of as the average size of the minority group. One finds (not shown) that it follows very closely the prevalence of dandies. Thus, allowing for originality somehow solves the paradox we started from: the wisdom of crowds can work while maintaining opinion diversity. 

In the labour division equilibrium, the population is neither a ``smear of ideas'' \cite{henrichCulturalGroupSelection2004a} nor homogenous. Instead, there is a clear majority, but also a substantial residual heterogeneity, which is endogenous and evolutionary stable. 

So, how can we position the agents in terms of conformism? Similarly to Fig.~\ref{fig:bias}, we can use our model (with $\omega=1$ so we fall in the labour division equilibrium) to plot the probability for a random individual to acquire an opinion by imitation, as function of opinion's prevalence in the population -- see Fig.~\ref{fig:bias}. We see that individuals are generally ``conformists'' in the sense of Boyd \& Richerson, unless the population is very close to full consensus, in which case the residual heterogeneity makes the curve saturate to a value $< 1$.  Thus, the population is bound to stabilize in one of the two stable fixed points (around $0.1$ and $0.9$), which means some opinion diversity is stable. Interestingly, Fig.~\ref{fig:bias} is very close to the type of curve postulated by Schelling or Granovetter \cite{schelling1971dynamic,granovetterThresholdModelsCollective1978} in their models of social imitation -- models that posit some inherent heterogeneities, for a review see \cite{bouchaudCrisesCollectiveSocioeconomic2013}.

Intuitively, agents are only cautiously influenced by the majority: they aggregate information while remaining wary of herding. Also, we can note that $q$ (the followers' accuracy, not shown) is a strictly increasing function of  opinion diversity, in line with Pages's ``Diversity Prediction Theorem'' \cite{pageDifferenceHowPower2008}: {\it the amount by which the crowd outpredicts its average member increases as the crowd becomes more diverse}. As noted in \cite{michard2005theory,bouchaudCrisesCollectiveSocioeconomic2013} such a situation also prevents large opinion swings (or ``crashes'') when external conditions slowly evolve. 

Note however that although the average value of $q$ is larger than $p$, we are still in the bi-stable region of Fig. \ref{fig:1}: in some instances, followers self-trap in the wrong belief. In our scenario, wisdom of crowd indeed exists but can sometimes badly fail. When neighbourhoods are not drawn at random in the population but according to static criteria (location, political opinion, social class, etc.) one should expect the formation of ``echo chambers'' with some clusters adopting the wrong belief, while others fully benefit from information aggregation.


\subsection{What about anti-conformism?}
\label{sec:anti-conformists}

\subsubsection{Is originality \textit{per se} a {credible} signal?}

The line of arguments developed above still has a loophole. If originality in itself is used by agents to detect the sources of genuine information, then original agents benefit from this signal by getting prestige in return. Thus, a social learner could benefit from being artificially original: he/she would be considered by others as a good source without paying the price of information search -- just like in Batesian mimicry, where the palatable species gains protection from predators without paying the cost of being toxic~\cite{wicklerMimicryEvolutionAnimal1965}. Furthermore, originality is simple enough to produce: an agent can choose opinions randomly, or sample around and adopt the \emph{minority} opinion.
Therefore, the above scenario is possibly vulnerable to an invasion by parasitic behaviour mimicking originality, thus scrounging the prestige of genuinely informed agents.

\begin{figure}
    \centering
    \includegraphics[width = 1\columnwidth]{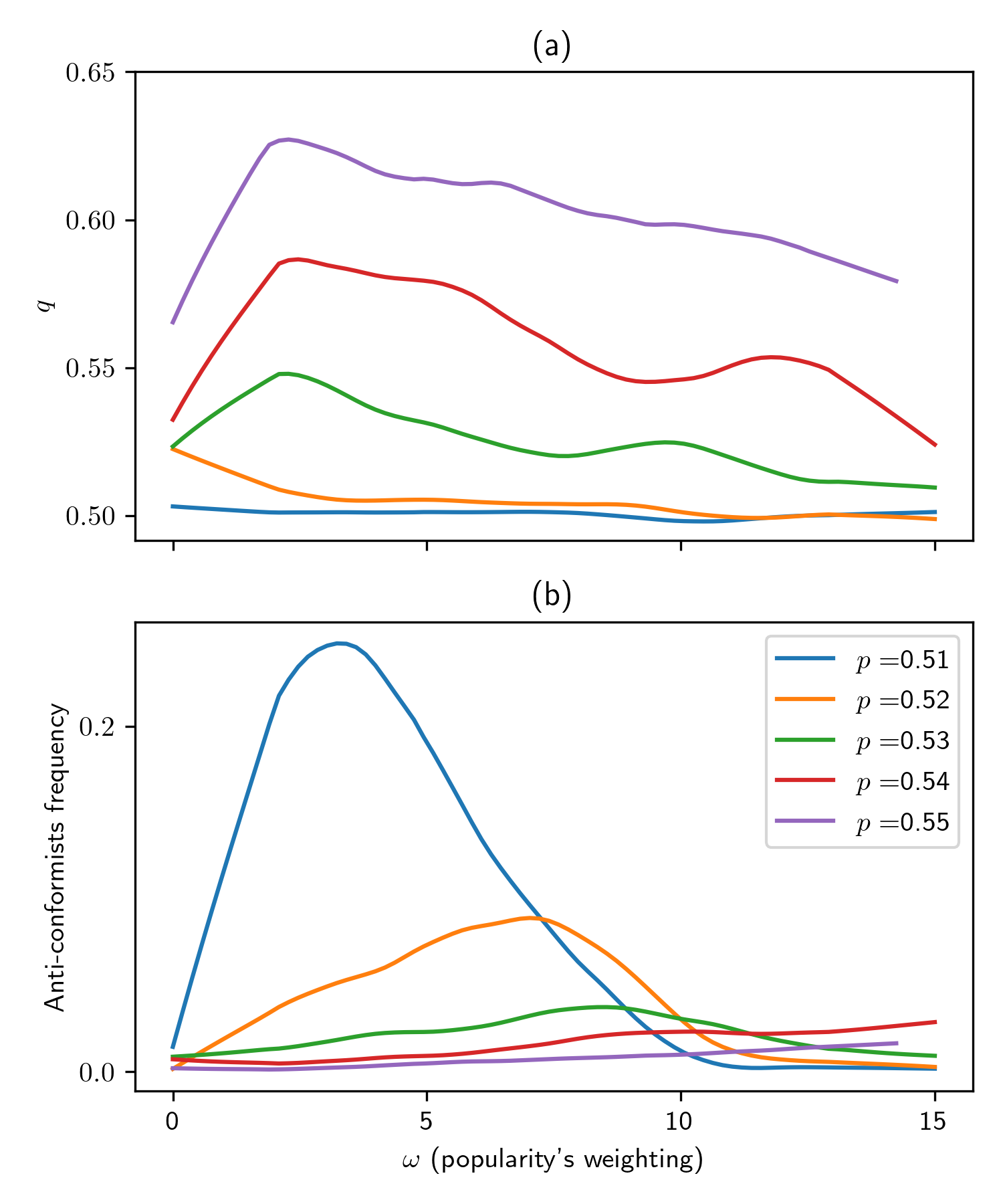}
    \caption{Effect of $\omega$ on $F$ success rate $q$ (a), and on anti-conformists prevalence (b), with $c=.01$ and various values of $p$. Lines are local regressions through data points (not shown for clarity).}
    \label{fig:omega_q2}
\end{figure}

It thus makes sense to include a fourth strategy: \emph{anti-conformism}, consisting in sampling around, and endorsing the \emph{minority} opinion. If such a behaviour spreads in the population, it challenges the very credibility of our scenario, as an original agent is no longer necessarily a source of reliable information.
To do so, we include a third binary gene, whose possession of the positive allele coupled with the follower one leads to anti-conformism. Like the other genes, it evolves by mutation and selection. Now, if we re-run the simulations of Fig.~\ref{fig:types}(a) allowing for this new strategy (see yellow curve in Fig.~\ref{fig:omega_q2}(a)), the $q>p$ phase disappears (in other words, the wisdom of crowds does not work anymore). 
The originality is at first used by dandies to hedge against herding, but is then exploited by anti-conformists, who become parasitic to the prestige of the informed. 

\subsubsection{The accuracy/prestige sweet-spot}

At this point, one could conclude that our scenario producing $q>p$ is flawed, and that it can only be saved if there exists an authentication method allowing truly informed agents to stand out. The situation is in fact more complex, in an interesting way.

Actually, anti-conformism yields originality at the expense of accuracy: if the population is better than random ($q>1/2$), then an anti-conformist mutant would be \emph{worse than random} ($1-q<1/2$). More mundanely, an anti-conformist's opinion is most certainly original, but is often balderdash. Making wrong forecasts decreases fitness in two ways: directly (Eq. \ref{eq:fitness}) and through prestige, as supplying good information is also key to obtaining prestige (recall the update rule \ref{eq:update_rule}).

The magnitude of this effect is controlled by the probability $p$ that an informed agent makes the right forecast (which is, contrarily to $q$, exogenous). We can think of $p$ as the easiness to forecast the topic in question. For instance, $p$ would be higher in meteorology than in finance, as the predictive power is (nowadays) stronger in the former case. So, it would make sense that the lower the $p$, the more anti-conformism can spread in the population. Put differently, the more unpredictable a subject is, the harder it is to detect con artists. Thus, the virtuous equilibrium described in Section~\ref{sec:results_1} is reachable when two conditions are met: (i) $p$ is sufficiently high to detect counterfeits; (ii) $\omega$ takes intermediate values, as in the previous section. 

This scenario is indeed confirmed numerically (cf. Fig. \ref{fig:omega_q2}(a)): the $q>p$ phase reappears for higher values of $p$. As shown in Fig.~\ref{fig:omega_q2}(b), anti-conformists only spread for low $p$ and intermediate $\omega$, in which case the division of labour fails.

In terms of opinion diversity (not shown), the phase where anti-conformists are present not only produces heterogeneity, but also \emph{polarisation}: the average size of the minority group is close to 50\%, i.e. the population is evenly split, as would indeed be expected for an anti-conformist bias (see Fig.~\ref{fig:bias}): if the probability to follow the minority is greater than the size of such minority, then the only stable fixed point corresponds to 50\% of the population holding one opinion, and 50\% holding the other.

\section*{Discussion}
\addcontentsline{toc}{section}{Discussion}

\subsection{Paradoxes of social transmission}

We started this paper with an apparent paradox. Condorcet's theorem \cite{condorcet1785essay}, together with empirical data \cite{novaestumpIndividualsFailReap2018,mesoudiExperimentalComparisonHuman2011}, shows that aggregating others' opinions often allows one to form more accurate opinions than relying on individual learning. If this is the case, individuals should rely on ``wisdom of crowds'' heuristics, and there is indeed evidence that they do \cite{mercierMajorityRulesHow2019}. However, as agents form their opinions based on the views of others, the diversity of opinions in the group diminishes. This undermines a necessary condition ensuring the very effectiveness of the wisdom of crowds in the first place, namely the independent formation of aggregated opinions. 

As a result, and as previous models suggest, we should end up in an apparently paradoxical equilibrium in which (i) the vast majority of agents adopt such a social learning strategy, leading to a large within-group homogeneity of opinions \cite{curtyPhaseCoexistenceForecasting2006,henrichEvolutionConformistTransmission1998}, even though (ii) this strategy, based on the wisdom of crowds effect, is not more effective than individual learning \cite{curtyPhaseCoexistenceForecasting2006}.  


Our model offers a way to solve this apparent paradox and generate endogenous heterogeneity. First, because listening to diverse opinions increases the efficiency of information aggregation \cite{pageDifferenceHowPower2008}, it is beneficial for individuals to listen preferentially to {\it original} opinions. Second, by assuming that individuals benefit not only from being well informed, but also from being listened to, the preference of social learners for original opinions endogenously creates an incentive to be independent-minded. 

As a result, as individuals have an extra-epistemic incentive to be listened to, some individual learning can endure despite the more efficient character of information aggregation. Importantly, the very persistence of individual learning guarantees the efficiency of social learning, as it introduces some diversity and independence in the opinions that the social learners gather -- a necessary condition for the wisdom of crowds to work.

\paragraph{Rogers' paradox.}
\label{rogers}

The paradox of the wisdom of crowd shares some similarities with Rogers' paradox \cite{rogersDoesBiologyConstrain1988}, and is thus relevant to its resolution as well. In Rogers' influential model, social learning is initially more efficient because it allows, in a varying environment, to avoid to cost of trial-and-error exploration. However, as social learners massively spread in the population, genuine information search get rarer. Social learning thus becomes less and less efficient, as the probability to copy an outdated belief or behaviour increases. Because of this negative prevalence dependence, social learning's fitness then collapses until being, at equilibrium, equal to the one of individual learning. Consequently, the mean fitness in such a group ends up being equal to what it would be in an ``acultural'' population, where everyone learns individually. In other words, culture (defined here as the ability to socially learn) would be, in the end, inevitable but useless. When compared to apparently massive contribution of social transmission to the ecological success of the human species \cite{henrichSecretOurSuccess2015,boydCulturalNicheWhy2011}, this result seemed somewhat paradoxical.

Several sophisticated solutions were proposed to solve Rogers' paradox, mostly involving hybrid strategies. Boyd \& Richerson's \cite{boydWhyDoesCulture1995} proposed ``selective learning'', where individuals try to learn individually first, then learn socially if the first trial was inconclusive. It allows to imitate opportunistically, only when needed, and predicts a fitness gain for the group at equilibrium. Enquist et al. \cite{enquistCriticalSocialLearning2007} proposed the reverse: ``critical learning'', consisting in learning socially first, then individually if the socially acquired strategy proved unsatisfactory.

Our model provides a simpler alternative to solve the conundrum. Under the hood, Rogers' result comes from the prevalence-independence of the fitness of individual learning strategies. Since by definition strategies have identical fitness at equilibrium, the system is bound to have the same fitness as an asocial learners population. In Boyd \& Richerson's model, this issue is bypassed by the assumption that the strategy consists in setting a threshold of self-confidence under which one learns socially, which means that the less one learns a-socially, the more one is demanding, hence the better its outcome. In our model, the positive prevalence-dependence of individual learners' payoff is quite natural: their opinions are sought for, so the rarer they are, the more they attract an audience.

\subsection{Predictions}
In addition to offering a solution to the above paradoxes, our model generates several predictions allowing one to explain some already known phenomena and possibly stimulate further empirical research. 

\paragraph{Egocentric discounting and the under-use of social information.}
Social learning seems to contribute to our species ecological success, and many models of its evolution showed its great adaptive benefits \cite{muthukrishna2016and, boydCultureEvolutionaryProcess1985}. However, somewhat unexpectedly, experimental evidence from social psychology, cultural evolution and experimental economics consistently shows that humans typically \textit{under-use} social information, by placing an unjustified weight on their own opinions relative to others' (see \cite{morin2020social}, for a comprehensive review).

This sub-optimal under-reliance on social information -- often called ``egocentric discounting'' \cite{yanivAdviceTakingDecision2000} -- is not easily explained within existing frameworks (see \cite{morin2020social}). By contrast, it is a natural consequence of our model. As social learners need to diversify their pool of opinion, individual learners, although informationally worse off, get social benefits from being listened to, that out-weight the costs of belief inaccuracy. In other words, a sub-optimal reliance on social learning with respect to belief accuracy is compensated by extra-epistemic benefits brought by social influence. 

In this perspective, egocentric discounting is explained by a mutually beneficial equilibrium, in which individual learners exchange, with social learners, original information for status. Such a division of labour equilibrium is consistent with the documented substantial inter-individual variation in the propensity to individually vs. socially learn \cite{morin2020social, olsen2019knowing, toelch2014individual}.

The same model also predict that (i) social learning is assorted to a taste for originality rewarding the independent-minded individual learners, and that (ii) individual learners, in exchange, gain social status or reputation benefits from their independent-mindedness. The next sub-sections successively examine these two points.

\paragraph{Minority influence and the diversification of informational sources.}
A taste for originality, by which people would put more weight on independent-minded individuals' opinions, has, to our knowledge, never been tested -- and could form the basis of future studies.

A general line of evidence on ``minority influence'' \cite{moscovici_1985} might however be a manifestation of such a phenomenon. In opposition to what a pure conformist bias would predict \cite{henrichEvolutionConformistTransmission1998}, minorities have indeed frequently been found to have a disproportionate social influence relative to their size in the population \cite{lataneSocialImpactMajorities1981,tanford1984social,erikssonArePeopleReally2009}. Again, this kind of phenomena emerge straightforwardly from the model: the taste for originality makes agents holding frequently deviant opinions more influential, and, as a result, the influence of a rare opinion is higher than its share in the population (cf Fig. \ref{fig:bias}). In this perspective, the disproportionate social influence of minorities would emerge from social learners' disposition to diversify their informational source to increase social learning's efficiency -- an adaptive form of bet-hedging.

\paragraph{Reputational benefits from information transmission.}
Moreover, for the ``division of labour'' to be mutually beneficial, this informational influence of independent minded people has to translate in terms of fitness benefits. Though social influence seems empirically related to fitness \cite{vonruedenWhyMenSeek2011}, it is important to discuss the possible reasons why this could be the case. We see two plausible foundations for this link. 

First, one could imagine an analogy to prestige biased social learning theory \cite{henrichEvolutionPrestigeFreely2001b} (though the analogy is somewhat misleading). This line of reasoning posits that, when some individuals have better-than-average skills in some domain (hunting, for instance), others individuals will pay deference to gain access to them, in order to imitate the best available models. It is akin to the way the informed get prestige in our model, by centralizing attention in the network. But at variance with our model the informed {\emph{are}} in this case informationally better off. In our virtuous equilibrium, informed agents are in fact the worst forecasters as they do not benefit from the Condorcet effect. But although their opinion is less accurate, it is paradoxically more informative about the state of the world, as it is not correlated to the opinion of the crowd. This extends Henrich \& Gil-White's logic: when sampling several models, it can make sense to pay a deferential cost even to someone who is not particularly competent, as long as his/her opinion is original.

Second, on a psychological level, our model can also be framed in terms of relevance theory \cite{sperberPrecisRelevanceCommunication1987a}. ``Originality'' is actually a way to characterize a ``relevant'' utterance in the sense of Sperber \& Wilson, which is, in turn, rewarded \cite{altay2020relevance}. Intuitively, Pinker \cite{pinkerLanguageInstinctHow1995} notes that {\it when a  dog bites a man that is not news, but when a man bites a dog that is news}. Formally, deviant opinion brings information in the Shannon sense, as it is considered as \textit{a priori} surprising. In pragmatic terms, an original statement is particularly relevant because there are many inferences to be drawn from it (for instance, {\it this guy does not follow the consensus, he might know something I don't}). In our model, agents develop endogenously a taste for original opinions. Somehow, the population learn spontaneously to treat such opinions as more relevant and to be wary of group-think.

In another formulation of the relevance theory, Dessalles \cite{dessallesAltruismStatusOrigin1998} notes that when communication amounts to sharing relevant information, it has an altruistic component. Since being relevant in a conversation is the rule rather than the exception, it requires an evolutionary explanation. Furthermore, Dessalles remarks that people often have a hard time being heard and making their point instead of being asked for information, which would be expected for an altruistic act. Humans seem to be willing not only to share information, but to strive to make others hear their point of view. A way to solve this paradox is to acknowledge that being listened and recognized as relevant, one builds a \emph{social status}. Similarly, when $\omega>0$ our model rewards agents according to their in-degrees, that is, the number of people listening to them. 

To put it more intuitively, Dessalles posits that {\it communication of information is part of a kind of unconscious trade in which status is the payment}, which is a good description of what happens in our model: the information cost paid is offset by the prestige brought by a large in-degree in the network. In some sense, the producer-scrounger dilemma that caused Rogers' paradox is solved by reputation mechanisms, which stabilizes a cooperative equilibrium (individual learning generates positive externalities). In line with this idea, prosocial individuals tend to spend more effort on individual learning \cite{erikssonBiasesAcquiringInformation2009}.

\subsection{Relevance to Efficient Market Theory}

Finally, as noted in the introduction, our story shares some similarities with, and offers some original insights on, the ``impossibility of efficient markets'' paradox discussed by Grossman \& Stiglitz in their seminal work \cite{grossmanImpossibilityInformationallyEfficient1980}. In particular, our distinction between information seekers (analysts), followers and anti-conformists is a relatively faithful description of the ecology of financial markets. (Note that some analysts may in fact themselves follow the consensus, see \cite{guedjExpertsEarningForecasts2004}).

Financial markets are notoriously difficult to predict, meaning, in our language, that the probability of success $p$ of the informed is very close to $1/2$. As we have pointed out in Section \ref{sec:anti-conformists}, this is a situation where ``quacks'' easily proliferate, since it is hard to distinguish them from genuine experts. In this case, the aggregation mechanism by which financial markets are supposed to reveal the ``true'' fundamental price of firms may completely breakdown. Furthermore, a prevalence of followers that trade chronologically {\it after} informed traders can give rise to bubbles that further decouple market prices from ground truth. 

Hence we are less optimistic than Grossman \& Stiglitz who propose that financial markets are close to efficiency, with a small residual error allowing information seekers to cover their costs while allowing the aggregation process to perform its task. We rather believe in Black's picture, where market prices are {\it within a factor of 2} of fundamental prices \cite{blackNoise1986,BlackWasRight2018}: only when mispricing becomes large enough will $p$ differ sufficiently from $1/2$, allowing true information to reveal itself and reinstate the power of Condorcet's aggregation theorem. In fact, since $p$ becomes itself endogenous in this case, financial markets are presumably prone to predictivity cycles: $p$ close to $1/2$ disfavours informed trading, which generates mispricing and, in turn, larger values of $p$, favouring informed trading, etc. It would be interesting to look for empirical signatures of such cycles. 

\subsection{Limitations}
We should nevertheless highlight a number of limitations of our model, which might help to delineate its domain of validity. For example, our model treats beliefs as perishable instead of cumulative: agents start searching from scratch at each iteration instead of taking their previous opinion as a starting point, and search to improve it. Also, the model implicitly assumes that agents can observe the opinions of their neighbours, but not assess directly the \emph{quality} of these opinions. At most, agents can estimate the quality of an informant through its accuracy record. Thus, our model is clearly unsuited to study a cultural item like canoe building techniques, which are cumulative and where the quality of a technique can be inferred through the quality of the produced canoes. This allows more elabourate strategies, such as the \emph{critical learning} \cite{enquistCriticalSocialLearning2007} we mentioned, or a \emph{prestige bias} \cite{henrichEvolutionPrestigeFreely2001b} (agents learning the best technique by imitating the most successful individual in the group). As these were not central to the contribution of the model, we chose not to include competence heterogeneity and prestige-biased transmission, although these are clearly major forces in cultural evolution. Our model is arguably more suited to analyse beliefs such as \emph{forecasts}. Indeed, it is impossible to compare the qualities of different forecasts before the event realization, and too late to change them afterward, which makes the two aforementioned solutions unworkable. 

Other mechanisms could also be invoked to justify the persistence of some individual learning (and the resulting preservation of some opinion diversity), despite social learning's efficiency. A possibility would be for more competent or knowledgeable individuals to discount the opinion of others in forming their own (see, e.g., \cite{hawthorne2019reasoning}). In line with this idea, studies suggests that IQ and self-confidence negatively predict reliance on social learning \cite{muthukrishna2016and, hawthorne2019reasoning}. This idea, however, assumes that it is because competent individuals are \textit{informationally better off} that they rely on individual learning (because they are ''wiser'' than the crowd). This, however, seems not to be the case. Throughout the experimental literature in social psychology, cultural evolution and experimental economics, individual learners' independent mindedness appears as informationally sub-optimal: they are found to \textit{under-use} social information relative to what an accuracy-maximization imperative would warrant \cite{morin2020social}. Furthermore, Mesoudi \cite{mesoudiExperimentalComparisonHuman2011} found that the subjects who used social learning the most were also above-average individual learners.

Finally, our model neglects possible conflicts of interest: information transmission has no reason to always be benevolent, but can lead to manipulation, against which individual learning is a way to hedge oneself. Also, our model does not factor in static biases in the linking process that can lead to ``echo chambers'', enhancing further the risk of manipulation, or of local failure of the wisdom of crowds. Recent social media examples abound, and in that respect one can only hope that individual learning remains prevalent. 

\begin{table}[b!]
    \centering
\begin{tabular}{c|l}
    $p$ & Informed accuracy \\ 
    \hline
    $q$ & Followers accuracy \\ 
    \hline
    $s$ & Fraction of links pointing toward informed agents\\
    \hline
     $\gamma$ & Memory decay rate \\
    \hline
    $\alpha$ & Taste for originality  \\
    \hline
    $\Omega_i$ & Network of  agent $i$ \\
    \hline
    $S_i^j$ & Score attributed by $i$ to $j$ \\
    \hline
    $\omega$ & Incentive to gather in-degrees  \\
    \hline
    $c$ & Information searching fitness cost \\
    \hline
    $\sigma$ & Mutation rate\\
\end{tabular}
\caption{}
\bigskip
\bigskip
\bigskip
\end{table}

\acknow{
We thank Daniel Nettle, Hugo Mercier, and Alexandre Darmon for fruitful discussions. This research was conducted within the \emph{Econophysics \& Complex Systems Research Chair}, under the aegis of the Fondation du Risque, the Ecole Polytechnique and Capital Fund Management.
}
\showacknow{}

\section*{Research Transparency and Reproducibility}
The Python code of the model is available \href{https://github.com/regicid/Condorcet_project}{here}.

\clearpage


\bibliography{Networks}

\end{document}